\documentclass[12pt]{iopart}

\usepackage{graphicx}
\usepackage{subfigure}
\include{eps}
\include{epsf}

\begin{document}

\title[Imprecise control and effects]{Imprecise k-space sampling and central brightening}

\author{R A Hanel, S De Backer, J Sijbers and P Scheunders}

\address{
Dept.\ of Physics, Vision Lab,   
University of Antwerp (CDE),   
Universiteitsplein 1,  
B-2610 Wilrijk, Belgium 
}

\ead{Rudolf.Hanel@ua.ac.be}

\begin{abstract}
In real-world sampling of k-space data, one generally makes a stochastic error not only in the value of the sample but in the effective position of the drawn sample. We refer to the latter as imprecise sampling and apply this concept to the fourier-based acquisition of magnetic resonance data. The analysis shows that the effect of such imprecisely sampled data accounts for contributions to noise, blurring, and intensity-bias in the image. Under general circumstances, the blur and the bias may depend on the scanned specimen itself. We show that for gaussian distributed imprecision of k-vector samples the resulting intensity inhomogeneity can be explicitly computed. The presented mechanism of imprecise k-space sampling (IKS) provides a complementary explanation for the phenomenon of central brightening in high-field magnetic resonance imaging. In computed experiments, we demonstrate the adequacy of the IKS effect for explaining central brightening. Furthermore, the experiments show that basic properties of IKS can in principle be inferred from real MRI data by the analysis on the basis of bias fields in magnitude images and information contained in the phase-images.       

\end{abstract}

\pacs{0250Ey, 4665, 7660Pc, 8385Fg}
\vspace{2pc}
\submitto{\PMB}

\section{Introduction}\label{section_introduction}
In this paper, a concept we would like to refer to as \textit{imprecise sampling} is introduced - and applied to MR image acquisition experiments. We will show that imprecise sampling provides an alternative explanation for the mechanism of \textit{central brightening}. This may be of interest in cases where the common interpretation in terms of dielectric resonance  
%
(e.g. Bomsdorf \etal 1988)
may fail. That may happen when high tissue conductivity prevents the creation of strong dielectric resonances 
%
(Collins \etal 2005).
The discussion adds to the recently revived discussion on this topic 
(Collins \etal 2005, Wiggins \etal 2005, Tropp 2004, Yang \etal 2002, Ibrahim \etal 2001, Roschmann 2000, Kangarlu \etal 1999). 
Nonetheless, the concept of imprecise sampling can as well be discussed in a self contained fashion. 
\par
Imprecise sampling basically relies on the observation that in real-world experimental situations sampling some property with respect to a continuous domain of control-parameters, such as sampling the k-space in an MRI experiment, can not be performed with infinite precision. We may intend to experimentally prepare and then sample the property for a specific value of the control parameter, i.e. ${\bf k}$, the index of the fourier-mode, but we will actually draw the sample somewhere near its intended value ${\bf k}$. This is, we will sample the property at some ${\bf k}+\delta {\bf k}$, where $\delta {\bf k}$ is a deviation from the precise k-space location ${\bf k}$, the \textit{imprecision}. Since standard Magnetic Resonance Imaging (MRI) is based on sampling the fourier-space of the underlying spin-relaxation distributions of the specimen with the k-space vectors ${\bf k}$ as control-parameter of the sampling process, the concept can directly be applied to this experimental situation. We will refer to this as \textit{imprecise k-space sampling} (IKS). 
\par
We have so far stated that imprecise control on the sampled fourier-mode inflicts deviations $\delta {\bf k}$ of the \textit{true} fourier-mode location ${\bf k}$ but we have not specified any properties of the inflicted deviations. Still, imprecision can be interpreted as some kind of noise corrupting the control-parameter. In the following, we will model the imprecision $\delta {\bf k}$ as a stochastic process and the properties of the imprecision can be expressed in terms of properties of the stochastic process. These properties have to be specified and argued. For instance, in a multi-echo sequence the fourier-modes are sampled along the read-direction by subsequent refocusing pulses and switching of read-gradients. The acquired samples will subsequently get more imprecise with each refocusing pulse and the cumulative error can be seen as a Wiener process. Similar arguments may hold in different sampling strategies as for instance spiral acquisition procedures. However, the contributions of this noise to the resulting image go beyond the additive noise model which is well known in the context of linear degradation models (LDM). 
\par
In many situations, image post-processing relies on the observation that there are several sources of degradation of an image that usually are classified into several categories, e.g. blur, noise, and intensity inhomogeneities (bias). These categories can be cast into degradation operators providing building blocks for LDMs which have successfully been utilized in image reconstruction and restoration methodology 
%
(Geman and Yang 1995,  Husse and Goussard 2004). 
However, when nonlinear effects of k-space imprecision become relevant, we can show that IKS provides a unified approach for explaining contributions to blurring and intensity inhomogeneities. 
\par
The theory of MR bias-fields is usually based on the analysis of the electro-magnetic field properties under specific geometric conditions and dielectric properties of the specimen in the scanner, e.g. 
%
(Hoult 2000). 
For known geometries, the Maxwell-equations can be computed and bias fields predicted, e.g. 
%
(Wang \etal 2002, Jin 1996). 
Usually, the precise imaging geometry is not available and image post-processing is confronted with the challenge of correcting intensity-inhomogeneities for instance for de-biasing, segmentation and tissue classification 
%
(Ahmed \etal 2002, Prastawa \etal 2005, Luo \etal 2005). 
All these explanations or predictions of bias-fields in general and \textit{central brightening} in particular argue very closely to the level of physics, i.e. electro-magnetism. 
\par
Central brightening is a well-known phenomenon in the context of high-field MRI 
%
(Collins \etal 2005, Tropp 2004). 
When the wavelength of the RF pulse used for slice selection, i.e. nuclear spin excitation, gets comparable with the dimension of the specimen the dielectric properties of the specimen, i.e. its tissue types, may influence the high frequency waves. This, for example, has been pointed out for cylindrical resonators by
%
Tropp (2004). 
Alternative explanations have been sought when dielectric resonance is questionable and, for instance, 
Collins \etal (2005) 
argues that high tissue conductivity prevents strong dielectric resonance. It is interesting to note that according to collected measurements of mammal tissue-conductivity 
%
(e.g. Gabriel \etal 1996)
changing the frequency domain from 127 MHz (3T) to 380 MHz (9T), white-matter renders an increase of tissue conductivity by a factor 2-3. However, central brightening is a sub-topic of the significant intensity inhomogeneities observed in high-field MRI 
%
(Vaughan \etal 2001, Bomsdorf \etal 1988) 
posing not only challenges for optimal coil design (e.g. 
%
Vaughan \etal 1994)
but also for the precise understanding of physical and physiological processes under the conditions of the given context.
\par
With IKS, we take a complementary approach to this phenomenon respecting the physical level, in terms of explaining the finite control on experimental control parameters, but dealing with the problem on the level of data acquired under conditions of imprecise control. We show later that for specific reasonable properties of the k-space imprecision $\delta {\bf k}$ the intensity-inhomogeneity from IKS can be computed explicitly and implies a gaussian shaped central brightening of the fourier-reconstructed images. The purpose of this paper is to present and 
demonstrate the concept of IKS. 
\par
The paper is organized as follows. We begin by introducing the IKS concept with respect to fourier-encoding MRI and how the imprecision can be seen as noisy control-parameters ${\bf k}$. We continue by showing how k-vector imprecision can be made responsible for contributions to visible blur, bias, and noise. 
We present an example that can be computed explicitly. Moreover, this example reproduces bias-fields with the characteristics of central brightening. We close by presenting characteristic computed images of IKS and discuss the results. 
\section{Fourier-encoded MRI and noise}\label{section_fourier_MRI}
In order to apply the idea of IKS to MRI, we recall that fourier-encoding is achieved by imposing phase-differences on the excited spins rotating at Larmor frequency by switching on linear magnetic gradients ${\bf B}_{i}(x)=g_i{\bf x}_i$, where $g_i$ are the constant gradient strength values, in ($i=0$) phase- and ($i=1$) read-direction. Without going into details, we can say that the phase-differences $\delta\phi_i \propto {\bf B}_i$ are proportional to the magnetic field strength in the respective directions. The trick of fourier-encoding is  to choose all effectively involved parameters such that the phase-shifts encode a sample in k-space, i.e. $\sum_i \delta\phi_i={\bf k}{\bf x}$ where ${\bf k}$ is the ideal fourier-vector, ${\bf x}$ is indexing the field of view (FOV). 
\par
The k-space vectors ${\bf k}$ are indeed control parameters of an MR imaging experiment. It is obvious that we can not achieve control with infinite precision. First of all, an MR scanner is a real device with finite accuracy in triggering read-out times and switching phase- and frequency-encoding gradients. Secondly, tissue may have different properties of interacting with the magnetic field related to the dielectric constant or the conductivity of the media. Now, suppose that the dielectric resonance is weak due to high tissue conductivity. Then, switching magnetic fields will induce stronger (ionic) currents in the tissue which adds to the effective magnetic field and locally distorts the magnetic field strength controlling the relative phase-shifts of the rotating nuclear spins. Yet, exactly these phase-shifts are preparing the experimental fourier-vector associated with the sampled fourier-mode. This tells us that the overall effect of these influences can be cast into an imprecision of the control-parameter, i.e. ${\bf k}\to{\bf k}+\delta{\bf k}$, with the effect that we do not measure the ideal fourier-modes of the sample-distribution but some effective fourier-mode 
\begin{equation}
\tilde{\rho}_{eff}({\bf k})=\int d{\bf x}\,\, \exp\left(i({\bf k}+\delta {\bf k}){\bf x}\right)\rho({\bf x})\quad ,
\label{IKS_modes}
\end{equation}
where $\rho({\bf x})$ is the ideal, static, spatial distribution of the spin-relaxation. 
\par
When we assume conditions where the interaction of the specimen and the magnetic field of the MR scanner lead to stochastic variations of the local magnetic field strength within the specimen, then the contributions to imprecision $\delta {\bf k}$ for a specific ${\bf k}$ have to be drawn from all over the specimen, i.e. for all locations ${\bf x}$, so that $\delta {\bf k}\equiv\delta {\bf k}({\bf k},{\bf x})$ is depending on both ${\bf k}$ and ${\bf x}$. In this sense, $\delta {\bf k}$ functionally depends on the spin-relaxation distribution $\rho$ of the specimen. Different tissue classes, accounting for possibly non-uniform spatial characteristics, may cause different local characteristics of imprecision for different values of the control-parameter ${\bf k}$.

\section{Bias, Blurring, and Noise}\label{section_degradations}

In this section, we discuss how k-vector imprecisions adds to degradations, i.e. to intensity inhomogeneities, blur, and noise, of the fourier-reconstructed images.
Let $\left<\right>$ denote the usual ensemble average or expectation-value with respect to the stochastic process $\delta {\bf k}$. 
Defining
\begin{equation}
G({\bf k},{\bf x})\equiv\left<\exp\left(i\delta {\bf k}{\bf x}\right)\right> \quad,
\label{G_funct}
\end{equation}
the expected effect of gradient noise on the fourier-basis vectors is 
\newline
$\left<\exp\left(i({\bf k}+\delta {\bf k}){\bf x}\right)\right>=G({\bf k},{\bf x})\exp\left(i{\bf k}{\bf x}\right)$. 
As a consequence, the average image of the simple fourier-reconstruction appears to be 
\begin{equation}
\tilde{\rho}_{*}({\bf k})=\int d{\bf x}\,\, \exp\left(i{\bf k}{\bf x}\right)G({\bf k},{\bf x})\rho({\bf x})\quad, 
\end{equation}
where $\tilde{\rho}_{*}\equiv \left<\tilde{\rho}_{eff}\right>$ is representing the expectation on an ensemble
of recorded data. Further, defining
\begin{equation}
g({\bf x}',{\bf x})\equiv\frac{1}{4\pi^2}\int d{\bf k}\,\, \exp\left(-i{\bf k}{\bf x}'\right)G({\bf k},{\bf x})\quad,
\label{g_fun}
\end{equation}
the inverse fourier-transform of the sampled fourier-data gives
\begin{equation}
\frac{1}{4\pi^2}\int d{\bf k}\,\, \exp\left(-i{\bf k}{\bf x}'\right)\tilde{\rho}_{*}({\bf k})=\int d{\bf x}\,\, g({\bf x}'-{\bf x},{\bf x})
\rho({\bf x})\quad.
\label{derived_IKS_formula}
\end{equation}
We see that $g({\bf x}'-{\bf x},{\bf x})$ has characteristics of a point-spread function via the ${\bf x}'-{\bf x}$ dependency with a local condition on ${\bf x}$ which refers to a bias. In case $\delta {\bf k}$ functionally depends on $\rho$, this dependency will be inherited by $g$ and the observable bias and blur will depend on the underlying nuclear spin relaxation distributions. 
\par
Note that, $g({\bf x}',{\bf x})$ basically can be a complex function. However, when $-\delta {\bf k}(-{\bf k},{\bf x})$ is equivalent to $\delta {\bf k}({\bf k},{\bf x})$ as a stochastic process then $g$ is a real function and no information should be found in the MR phase image. Moreover, when $\left<\delta {\bf k}_i^n\right>=0$ for all odd $n$, then $G$ is a real function and $g(-{\bf x}',{\bf x})^{*}=g({\bf x}',{\bf x})$, where $*$ denotes complex conjugation. 
\\
%
\par
We would like to note that the derivation of equation (\ref{derived_IKS_formula}) is based on the average data $\tilde{\rho}_*$.
Simulated single images with an imprecision $\delta {\bf k}$  sampled on the image grid size are extremely corrupted and it takes an average over many images 
to produce 
a reasonable signal to noise ratio (SNR). 
We might ask why it is possible to see the central brightening effect in single real recorded images after all? 
To see this, one should note that in the MRI experiment the fourier-modes are sampled one by one. Yet, each time we sample some specific ${\bf k}$, i.e. $\tilde{\rho}_{eff}({\bf k})$, the integral in equation (\ref{IKS_modes}) is collecting the stochastic events of imprecision over the entire spatial domain. This is comparable to the ensemble average in case the spatial correlation length of the stochastic process $\delta{\bf k}$ is small with respect to the voxel size of the image-grid. The spatial correlation length defines the voxel-size of a fine grid. Imprecision simulated on the fine grid can then be down-sampled to the coarser image grid. But down-sampling is equivalent with averaging over a large number of coarse-grid-images whith their coarse-voxel values drawn from the ensemble of fine-voxel values contained in the coarse-voxel. 
Finite but small spatial correlation lengths of $\delta{\bf k}$ comparable to the voxel-size, however, will not be completely smoothed by integration over the domain of a coarse-voxel and the events of imprecision remain visible after fourier-transforming back to x-space. This remaining imprecision is basically not distinguishable from additive noise while the contributions that have been effectively averaged by integration appear as blur and bias.  
\\
\\
\textit{Example}: Let us look at the stochastic process $\delta {\bf k}({\bf k},{\bf x})$ and assume the components $\delta {\bf k}_i$  to be normally distributed with zero mean and co-variance $D_{ij}({\bf k},{\bf x})=\left<\delta {\bf k}_{i}\delta {\bf k}_{j}\right>$. The standard deviation is denoted $\sigma$ with $\sigma^t\sigma=D$, where $\sigma^t$ denotes the transposed matrix of $\sigma$. 
Let us further assume point-wise statistical independence of the components $\delta {\bf k}_i$, i.e. 
$\left<\delta {\bf k}_{0}^m\delta {\bf k}_{1}^n\right>=\left<\delta {\bf k}_{0}^m\right>\left<\delta {\bf k}_{1}^n\right>$ for all 
$({\bf k},{\bf x})$. This can be justified since phase- and frequency-encoding gradients are generally not switched on at the same time and implies that $D$ and thus $\sigma$ are diagonal. Then it is possible to explicitly compute $G$ and $g$. Starting with the definition of $G$ in (\ref{G_funct}) we expand the exponential into a series, and use that the $(2n)$'th central moment of a gaussian process equals to $(\sigma^2/2)^n(2n)!/n!$, and all odd moments vanish. This implies that in the expansion 
\begin{equation}
G({\bf k},{\bf x})=\sum_{n=0}^{\infty}\frac{(i)^n}{(n)!}\sum_{m=0}^n{n \choose m}\left<\delta {\bf k}_0 {\bf x}_0\right>^{n-m}
\left<\delta {\bf k}_1 {\bf x}_1\right>^m
\end{equation}
only terms with even $n$ and even $m$ contribute.  After minor manipulations, one arrives at the following expression:
\begin{equation}
G({\bf k},{\bf x})=\exp\left(-\frac12 \left|\sigma({\bf k},{\bf x}){\bf x}\right|^2\right)\quad . 
\end{equation}
Suppose $\sigma$, or equivalently $G$, is not depending explicitly on ${\bf k}$, then the definition of $g$ in (\ref{g_fun}) implies that 
\begin{equation}
g({\bf x}'-{\bf x},{\bf x})=\delta({\bf x}'-{\bf x})G({\bf x})
\label{delta_g}
\end{equation}
and thus
\begin{equation}
\int d{\bf x}\,\, g({\bf x}'-{\bf x},{\bf x})\rho({\bf x})=G({\bf x}')\rho({\bf x}')\quad.
\end{equation}
In this case we can not only compute the bias field explicitly but also note how the bias field may functionally depend on $\rho$ via the 
${\bf x}$-dependency of $\sigma$, as different tissue types correspond to different spatial domains with different standard deviations of imprecision. When the imprecision is ${\bf k}$-independent no blurring is introduced in the image and we have pure bias. When $\sigma$ is not depending on ${\bf x}$ as well, i.e. there is no functional dependency on the ground truth, then we explicitly computed a gaussian intensity bias with the maximum in the image center. We want to point out how this simple example in particular demonstrates how IKS, as a general concept, provides a complementary explanation for the phenomenon of \textit{central brightening}. Moreover, we have to note that in the general case $\sigma\equiv\sigma({\bf k},{\bf x})$ the bias-point-spread-function $g$ can not be written as in (\ref{delta_g}) with the immediate consequence that noise levels $\sigma$ varying with ${\bf k}$ have to introduce blurring or distortions in the image of $\rho$.
\section{Experiments}\label{section_experiments}
\begin{figure}[hbtp]
	\begin{center}
	\mbox{ 
		\subfigure[Phantom]{		\includegraphics[width=0.29\columnwidth]{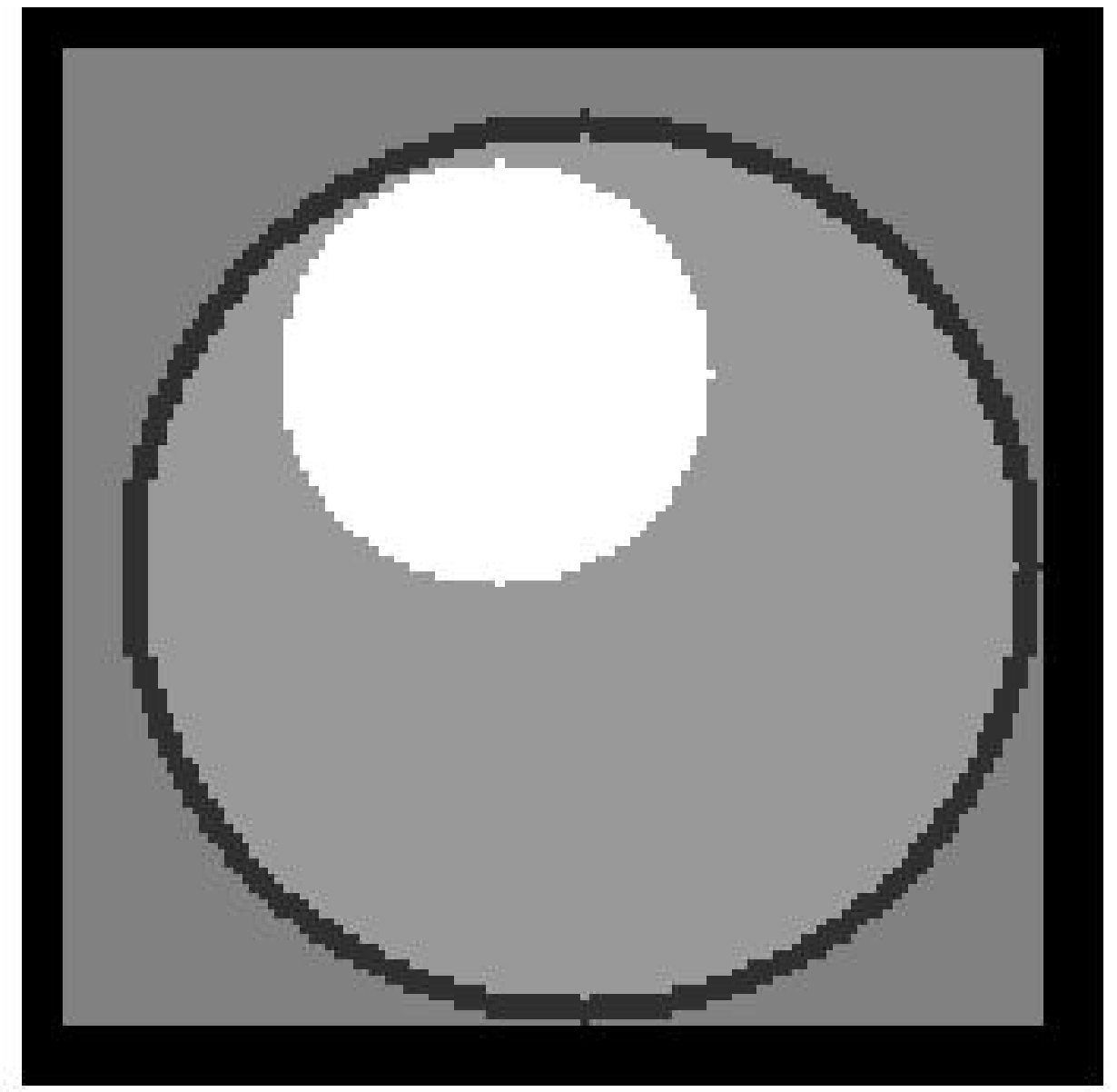}} \quad 
	}
	\mbox{ 
		\subfigure[Simulated]{		\includegraphics[width=0.29\columnwidth]{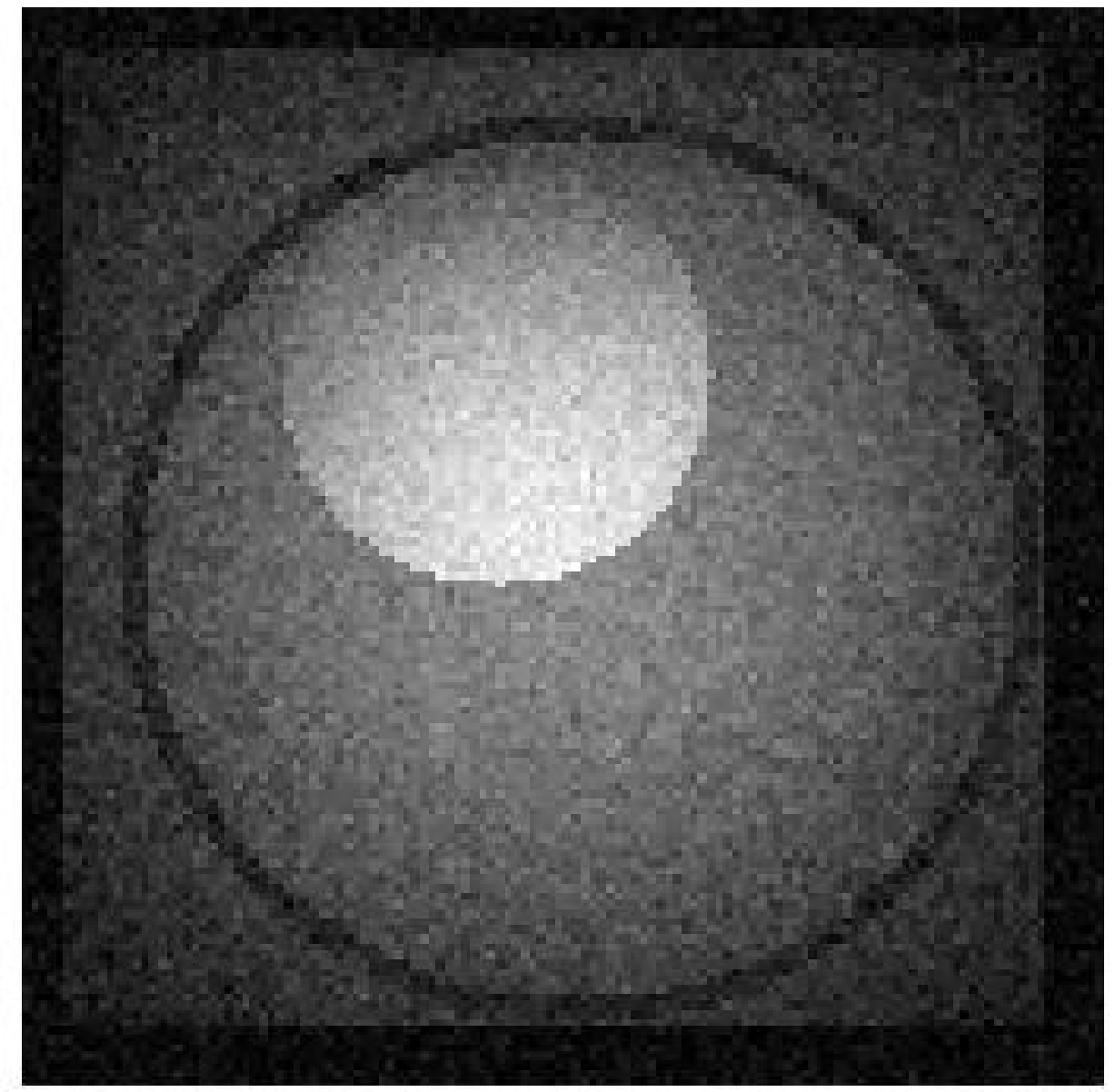}} \quad 
	}
	\mbox{ 
		\subfigure[Computed]{		\includegraphics[width=0.29\columnwidth]{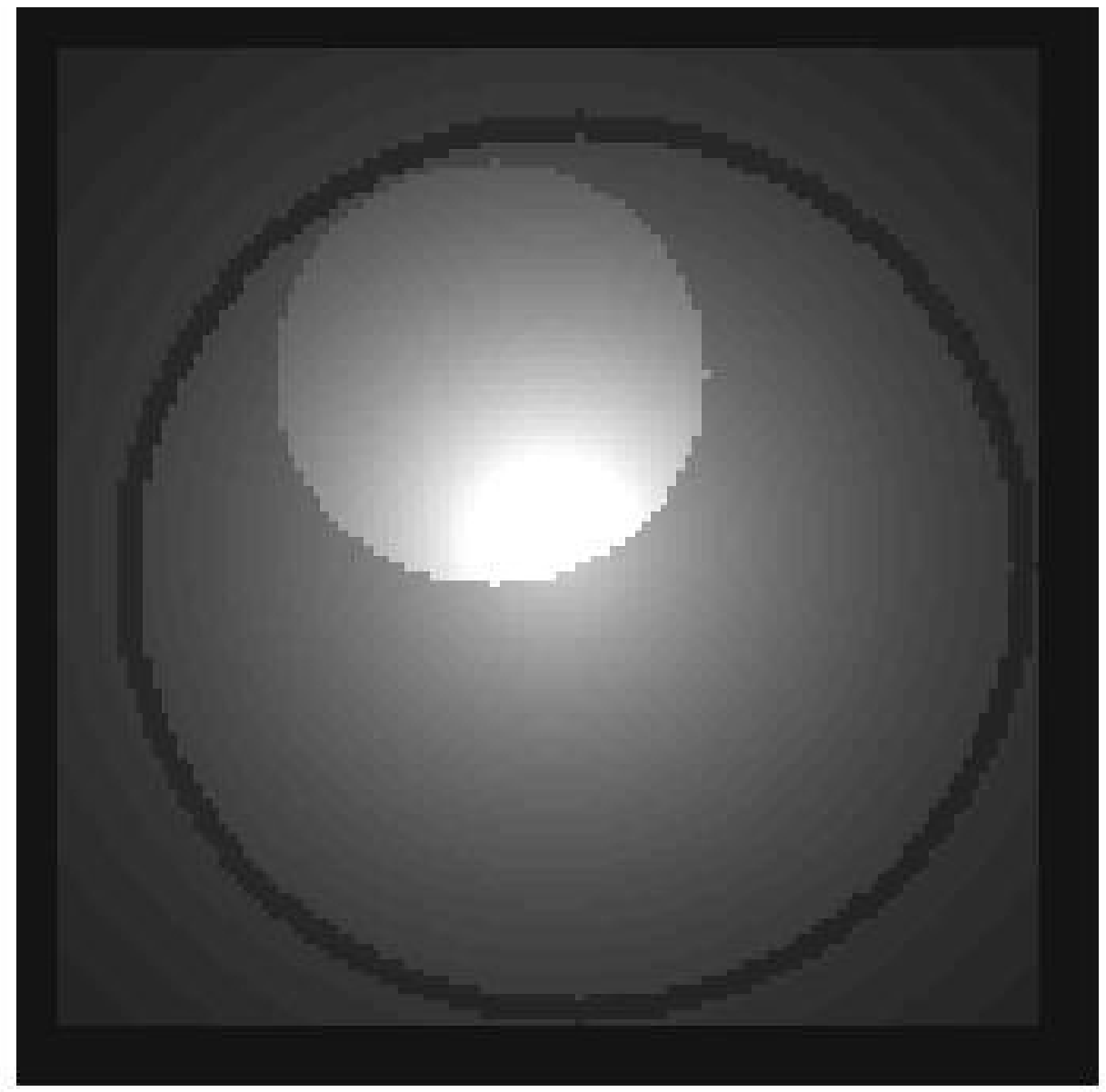}} \quad
	}
\caption{\small{Simulated and Computed IKS effect for constant $\sigma$.}}
\label{comp_vs_sim_img}
\end{center}
\end{figure}

One can alternatively compute the IKS effect according to equation\ \ref{derived_IKS_formula} or simulate the IKS effect. The simulations can be done by effectively computing fields $\delta {\bf k}({\bf k},{\bf x})$ for every ${\bf k}$, multiply the associated factor $\exp(i\delta{\bf k}{\bf x})$  point-wise with the phantom-distribution $\rho$ perform a fast-fourier-transformation (FFT), and only choose one $k$-mode according to the sampling sequence. After all fourier-modes have been sampled in this manner we transform the sampled fourier-information back to x-space with an inverse FFT. Simulations are rather time intensive compared to computing images but demonstrate the contribution of IKS to noise, which is lost in the computed images. In figure (\ref{comp_vs_sim_img}, b,c) we show this for constant $\sigma$.  
%
%
%
%

Our experiments consist of computed and simulated effects of IKS on a known software-phantom $\rho({\bf x})$ (figure\ \ref{comp_vs_sim_img}, a). 
In order to get an impression of the IKS effect we look at solutions of the example given in section (\ref{section_degradations}) and choose some simple 
models for the standard deviation of the imprecision
\begin{equation}
\begin{array}{l}
\sigma_{I}({\bf x}) = \sigma_0\rho({\bf x})^{\alpha}\quad,\\
\sigma_{II}({\bf k})=\sigma_0(1+|{\bf k}|)^{\beta}\quad, \\
\sigma_{III}({\bf k})=\sigma_0\left(1+\gamma({\bf k}_0-k_{\mbox{off}})\right)\quad ,
\end{array}
\end{equation}
where $\sigma_0$ is an amplitude factor, $\alpha$, $\beta$, and $\gamma>0$ supply an additional parameter to each model. In the $\sigma_{III}$ model
$k_{\mbox{off}}$ is an offset such that for the first sample in read direction ($i=0$) $\delta{\bf k}$ has a standard deviation $\sigma_0$ and the standard deviation of the imprecision increases with $\sigma_0\gamma\Delta {\bf k}_0$ for any subsequent fourier-mode in read direction, where $\Delta {\bf k}_0$ is the increment between two subsequent fourier-modes in read direction.
In the first model we can associate different levels of imprecision with different regions of intensity-values of $\rho$ representing the different tissue types. While $\sigma_{I}$ models imprecision that may vary depending only on the origin of imprecision, i.e. its functional dependency on $\rho$, the models $\sigma_{II}$ and $\sigma_{III}$ are roughly modeling effects that can be associated with the methodology of sampling.
Particularly $\sigma_{III}$ can be regarded as a model for the linear increasing standard deviation of a Wiener-process in read direction.
%
\begin{figure}[hbtp]
	\begin{center}
	\mbox{ 
		\includegraphics[width=\columnwidth]{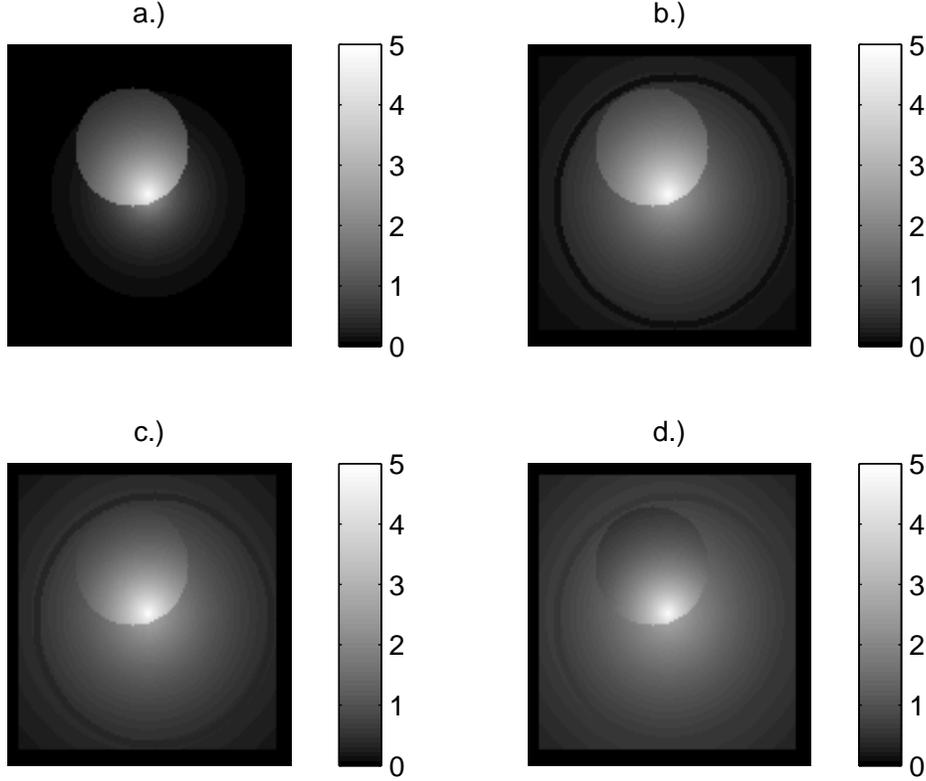}
	}
	\caption{\small{
		Model $\sigma_{I}$, with $\alpha=-1$ (a), $\alpha=-0.5$ (b), $\alpha=0.5$ (c), $\alpha=1$ (d); 
		The amplitude $\sigma_0$ has been chosen such that the smaller circular structure is rendered with comparable gray values.
	}}
\label{x_series_img}
	\end{center}
\end{figure}
\begin{figure}[hbtp]
	\begin{center}
	\mbox{ 
		\includegraphics[width=\columnwidth]{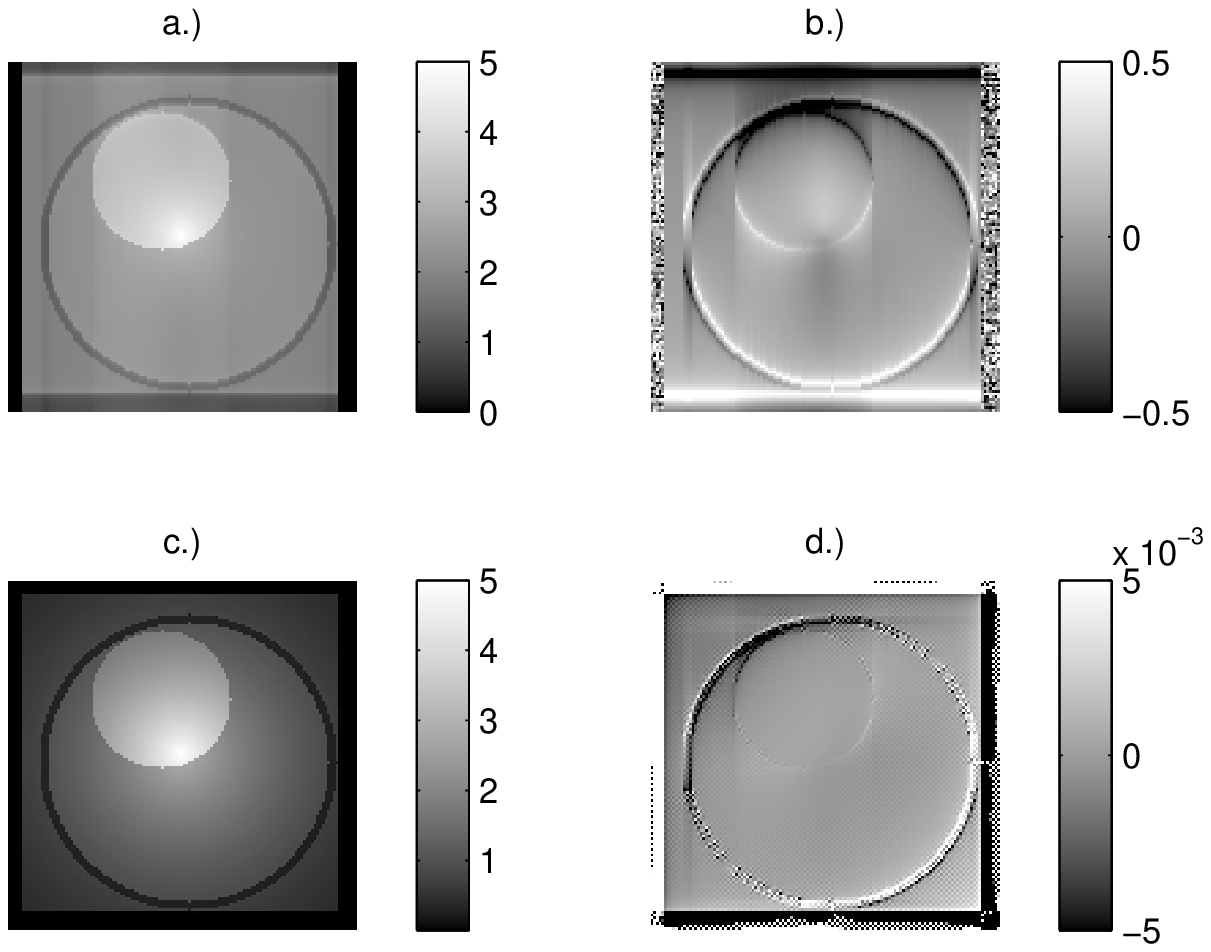}
		}
	\caption{\small{		
		(a) amplitude- and (b) phase-image of the radial model $\sigma_{II}(\beta=0.1$) \\
		(c) amplitude- and (d) phase-image of the multi-echo model $\sigma_{III}(\gamma=0.001$) \\		 
	}}
\label{k_series_img}
	\end{center}
\end{figure}
\\
We note that these choices have been made in order to visualize some basic properties of the IKS effect. Our choices are extremely simple models that still have the capacity to demonstrate what effects we may expect from IKS with only a few parameters. Moreover, since we work with simulations on simple software phantoms, we only have to discriminate a small number of different "tissues". A power of $\rho$ is therefore sufficient to explore the situation in x-space. In k-space, a power in $|{\bf k}|$ can roughly be associated with a spiral or radial sampling protocol, while the linear model $\sigma_{III}$ may be associated with the cumulative error made in preparing the fourier-modes in a multi-echo sequence in read-direction. 
\\
\par
The experiments show the following. First of all, that the simulated effect agrees well with the computed effect (figure\ \ref{comp_vs_sim_img}).
The simulated image is the average of 40 images with simulated imprecision. The images in figure\ (\ref{x_series_img}) are computed with the imprecision-model $\sigma_{I}$, which is not depending on ${\bf x}$. We can see how different choices of $\alpha$ associates different radial decay rates to the different "tissues", which demonstrates how the effect of IKS functionally depends on the specimen. Stronger tissue related imprecision causes a stronger radial decay of intensity in the tissues gray-values. Moreover, the imaginary part of these images vanishes up to computing accuracy so that one may conclude, that it is not the x-space related functional dependency of imprecision, that causes non-vanishing imaginary images. 
\par
The images in figure\ (\ref{k_series_img}), computed with the imprecision-model $\sigma_{II}$ and $\sigma_{III}$, show that ${\bf k}$-depending imprecision introduce non vanishing imaginary parts of the resulting images and therefore also non-vanishing phase-images (figure\ \ref{k_series_img}, b,d). The magnitude images (figure\ \ref{k_series_img}, a,c) show IKS for the radial and the spin-echo model and it is obvious that the radial imprecision introduces a much weaker phase-image although the amplitude images show comparable central brightening. Furthermore the procedural asymmetry between phase-encoding and read - direction in the spin-echo model can be noticed clearly. 
\par
In consequence it is, at least in principle, possible to study imprecision on the basis of real experiments. The problem can be separated into contributions to the IKS effect from x-space, regarding to central brightening decaying at different rates in different tissue types, and contributions from k-space, i.e. the sampling process itself, regarding to information contained in phase-images. 
One may also note the implication that under specific circumstances it should be possible to distinguish tissue with identical spin-relaxation properties T1 and T2 and thus basically have to be rendered with identical gray-values as long as they introduce different levels of imprecision.
In this hypothetical situation the tissue would become distinguishable under high-field conditions due to different radial decay-rates of the image intensity. Imprecision may therefore possibly provide information that is complementary to spin-relaxation and regards to properties like the dielectric constant and the tissue conductivity.

\section{Conclusions}\label{section_conclusions}
We have introduced the concept of \textit{imprecise k-space sampling} (IKS) and demonstrated, that this concept provides an alternative explanation for the phenomenon of central brightening in high field MRI. We have demonstrated by computer experiments, on the basis of simple imprecision-models, that (i) such imprecissions do have the effect the theory predicts, (ii) functional x-distribution dependency alters tissue-decay rates, while it can not be made responsible for non-vanishing imaginary-images, (iii) functional k-distribution dependency is capable of producing non-vanishing imaginary images. These results show, that it is possible to collect information on the imprecision from the complex fourier-data of an MRI experiment by analysis of intensity inhomogeneities and phase images. However, that imprecise controll has consequences beyond noise, is a finding as basic as surprising.  
%
\ack
The IBBT - ICA4DT project is co-funded by the IBBT (Interdisciplinary Institute for BroadBand Technology), a research institute founded by the Flemish Government in 2004, and the involved companies (Agfa, Barco, Medicim, Namahn). \\
https://ica4dt.ibbt.be

%
%
\References
%
\item[]
Ahmed M N, Yamany S M, Nevin M, Farag A A and Moriarty T
2002
A modified Fuzzy C-Means Algorithm for Bias Field Estimation and Segmentation of MRI Data
{\it IEEE Trans Med Imaging} {\bf 21} (3) 193--199
%
\item[]
Bomsdorf H, Helzel T, Kunz D, Roschmann P, Tschendel O and Wieland J
1988
Spectroscopy and Imaging with a 4 Tesla whole-body MR system
{\it NMR Biomed} {\bf 1} (3) 151--158
%
\item[]
Collins C M, Wanzhan Liu, Schreiber W, Qing X Yang and Smith M B 
2005
Central brightening due to constructive interference, with, without, and despite dielectric resonance
{\it J. Magnetic Resonance Imaging, Technical Note} {\bf 21} 192--196
%
\item[]
Gabriel C, Gabriely S, and Corthout E
1996
The dielectric properties of biological tissues: I. Literature survey
{\it Phys. Med. Biol.} {\bf 41} 2231–-2249
%
\item[]
Geman D and Yang C
1995
Nonlinear Image Recovery with Half-Quadratic Regularization
{\it IEEE Transactions on Image Processing} {\it 4} (7) 932--946 
%
\item[]
Hoult D I
2000
The principles of reciprocity in signal strength calculations; a mathematical guide
{\it Concepts in Magnetic Resonance} {\bf 12} 173--187
%
\item[]
Husse S and Goussard Y
2004
Extended forms of Geman \& Yang algorithm: application to MRI reconstruction
{\it Proc. IEEE ICASSP} {\bf 3} 513--516
%
\item[]
Ibrahim T S, Lee R, Abduljalil A M, Baertlein B A and Robitaille P M
2001
Dielectric resonances and B1 field inhomogeneity in UHFMRI: computational analysis and experimental findings 
{\it J. Magnetic Resonance Imaging} {\bf 19} 219–-226
%
\item[]
Jin J M
1996
Computation of electromagnetic fields for high-frequency magnetic resonance imaging applications
{\it Phys. Med. Biol} {\bf 41} 2719--2738
%
\item[]
Jinghua Wang, Qing X Yang, Xiaoliang Zhang, Collins M C, Smith M B, Xiao-Hong Zhu, Adriany G, Kamil Ugurbil and Wei Chen
2002
Polarization of the RF Field in Human Head at High Field: A Study with  Quadrature Surface Coil at 7.0 T
{\it Magnetic Resonance in Medicine} {\bf 48} 326--369
%
\item[]
Jianhua Luo, Yuemin Zhu, Clarysse P and Magnin I
2005
Correction of Bias Field in MR Images Using Singular Function Analysis
{\it IEEE Trans Med Imaging} {\it 24} (8) 1067--1085
%
\item[]
Kangarlu A, Baertlein B A, Lee R, Ibrahim T, Yang Lining, Abduljalil A M and Robitaille P-M L
1999
Dielectric resonance phenomena in ultra high field MRI 
{\it J Comp. Ass. Tomography} {\bf 23} 821–-831
%
\item[]
Prastawa M, Gilmore J H, Weili Lin and  Gerig G
2005
Automatic segmentation of MR images of the developing newborn brain
{\it Medical Image Analysis} {\bf 9} (5) 457--466
%
\item[]
Qing X Yang, Jinghua Wang, Xiaoliang Zhang, Collins C M, Smith M B , Haiying Liu, Xiao-Hong Zhu, Vaughan J T, Kamil Ugurbil and Wei Chen
2002 
Analysis of wave behavior in dielectric sample at high field 
{\it Magnetic Resonance Med} {\bf 47} 982–-989
%
\item[]
Roschmann P
2000 
Role of B1 eigenfields of dielectric objects in highfield MRI
{\it Proc. of the 8th Annual Meeting of ISMRM, Denver} 151
%
\item[]
Tropp J 
2004
Image brightening in samples of high dielectric constants
{\it J. Magnetic Resonance Imaging} {\bf 167} 12--24
%
\item[]
Vaughan J T, Garwood M, Collins C M, \etal
2001
7T vs. 4T: RF Power, Homogeneity, and Signal-to-Noise Comparison in Head Images
{\it Magnetic Resonance in Medicine} {\bf 46} 24--30
%
\item[]
Vaughan J T, Hetherington H P, Otu J O, Pan J W and Pohost G M
1994
High frequency volume coils for clinical NMR imaging and spectroscopy
{\it Magnetic Resonance in Medicine} {\bf 32} 206--218
%
\item[]
Wiggins G C, Potthast A, Triantafyllou C, Wiggins C J and Wald L L 
2005
Eight-channel phased array coil and detunable TEM volume coil for 7 T brain imaging
{\it Magnetic Resonance in Medicine} {\bf 54} 235–-240


\endrefs

\end{document}